\documentclass[prb,preprint,groupedaddress,showpacs]{revtex4}
\usepackage{epsfig}
\begin{document}

\title{Mg impurity in helium droplets}
\author{J. Navarro}
\affiliation{IFIC (CSIC-Universidad de Valencia), Apartado Postal 22085,
E-46.071-Valencia, Spain}
\author{D. Mateo}
\affiliation{Departament E.C.M., Facultat de F\'{\i}sica, and IN$^2$UB, 
Universitat de Barcelona. Diagonal 647, E-08028 Barcelona, Spain}
\author{M. Barranco}
\affiliation{Departament E.C.M., Facultat de F\'{\i}sica, and IN$^2$UB, 
Universitat de Barcelona. Diagonal 647, E-08028 Barcelona, Spain}
\author{A. Sarsa}
\affiliation{Departamento de F\'{\i}sica, Facultad de Ciencias, 
Universidad de C\'ordoba. E-14071 C\'ordoba, Spain}

\date{\today}

\begin{abstract}
Within the diffusion Monte Carlo approach,
we have determined the structure of isotopically pure
and  mixed helium droplets  doped with one magnesium atom.
For pure $^4$He clusters, our results confirm those of 
M. Mella {\it et al.} [J. Chem. Phys. {\bf 123}, 054328 (2005)]
that the impurity experiences a transition
from a surface to a bulk location as the number of helium atoms
in the droplet increases. 
Contrarily, for pure $^3$He clusters Mg 
resides in the bulk of the droplet due to the smaller surface
tension of this isotope. Results for mixed droplets are presented.
We have also obtained the absorption spectrum of Mg around the
$3s3p \, ^1P_1 \leftarrow 3s^2 \, ^1S_0 $ transition.

\end{abstract}



\maketitle

\section{Introduction}

The solubility of alkaline earth atoms attached to helium
droplets has been found to critically depend on the dopant and on the
helium isotope droplets are made of. This is at clear variance with what
happens for alkali atoms, that are all found to reside in a dimple at
the surface of the droplet irrespective of their isotopic
composition,\cite{Bue07,Mat11}
or to impurities that experience a large attractive interaction with
helium as, {\it e.g.}, inert gas atoms that are found to reside in the
bulk of the droplet. This is a particular yet prominent aspect of a
much broader subject -the physics and chemistry of pristine and
doped helium nanodroplets- that has
been reviewed in a series of articles, see {\it e.g.} Refs.
\onlinecite{Toe04,Bar06,Sti06,Cho06,Tig07,Sza08} and references
therein.

Ancilotto and coworkers\cite{Anc95} have provided a solvation criterion
for impurities in liquid helium in terms of the dimensionless parameter
\begin{equation}
\lambda = \frac{\rho \, \varepsilon \, r_e}{2^{1/6} \, \sigma} \, ,
\end{equation}
where $\rho$ and $\sigma$ are the density and the surface tension of liquid He,
respectively, 
and $\varepsilon$ and $r_e$ are the well depth and the equilibrium distance of
the He-impurity interaction, respectively.
This parameter measures the balance between the energy of the impurity and
the surface energy of the liquid. If $\lambda<1.9$, the impurity 
sits on the free surface of the fluid and no solvation occurs. For Mg 
one has  $\lambda=2.6$ for $^4$He and $\lambda=4.6$ for $^3$He.
Since this criterion does take into account neither the complexities 
of the system nor the fine details of the He-impurity pair potential
and it has been established for bulk liquid, it must be
taken with care when the value of $\lambda$ is fairly close to 1.9,
in which case only a detailed calculation may
unveil the solvation properties of a given impurity in He drops.

It has been found experimental and theoretically that Ca atoms solvate
in $^3$He but not in $^4$He droplets, and that depending on their
isotopic composition, they may reside in the $^3$He-$^4$He
interface that develops in mixed droplets at low
temperatures.\cite{Her07,Bue09,Gua09}
The lighter alkaline earth Mg presents a borderline behavior. Within
Density Functional Theory (DFT), it has
been found to solvate in $^3$He droplets and to be very delocalized in
$^4$He ones.\cite{Her07,Her08} Previous diffusion Monte Carlo
calculations (DMC)\cite{Mel05} have yielded the result
that Mg is fully solvated for a critical number of $^4$He atoms 
of about $N_4 = 30$. A similar transition from surface to bulk location
of Mg was also found within DFT.\cite{Her08} Experimentally,
full solvation of Mg in $^4$He droplets has been inferred from
the analysis of Laser Induced Fluorescence (LIF),\cite{Reh00}
comparing it with
LIF experiments on the absorption spectrum of Mg atoms in
liquid $^4$He.\cite{Mor99,Mor06} More recent
Resonant Two-Photon-Ionization (R2PI) experiments\cite{Prz08}
have also yielded a bulk location for this impurity.

Since an electron-impact ionization experiment\cite{Ren07}
carried out in Mg doped
$^4$He drops with about 10$^4$ atoms seems to indicate that magnesium
reside at or near the droplet surface, and some unpublished 
DMC calculations seem also to point toward a rather
surface location of Mg in $^4$He droplets,\cite{Elh07,Elh08,Elh09}  
we have undertaken an independent Monte Carlo analysis of the structure
of these systems aiming at settling this issue,
complementing the previous analyses with a  DMC
study of $^3$He and mixed $^3$He-$^4$He droplets doped with Mg that has
not been previously carried out.
The interest in addressing doped mixed helium clusters has been already
stressed.\cite{Bue07,Gua09} Very recently, we have studied the structure
and absorption spectrum of Mg in liquid helium mixtures as a function of
pressure and isotopic composition.\cite{Mat11}

This paper is organized as follows.
 In Sec.~\ref{method} we give some details about the DMC
calculations. In Sec.~\ref{results} we present our results, and in Sec.~\ref{conclusion}
we present a brief summary.

\section{Methods}
\label{method}

\begin{figure}[ht]
\epsfig{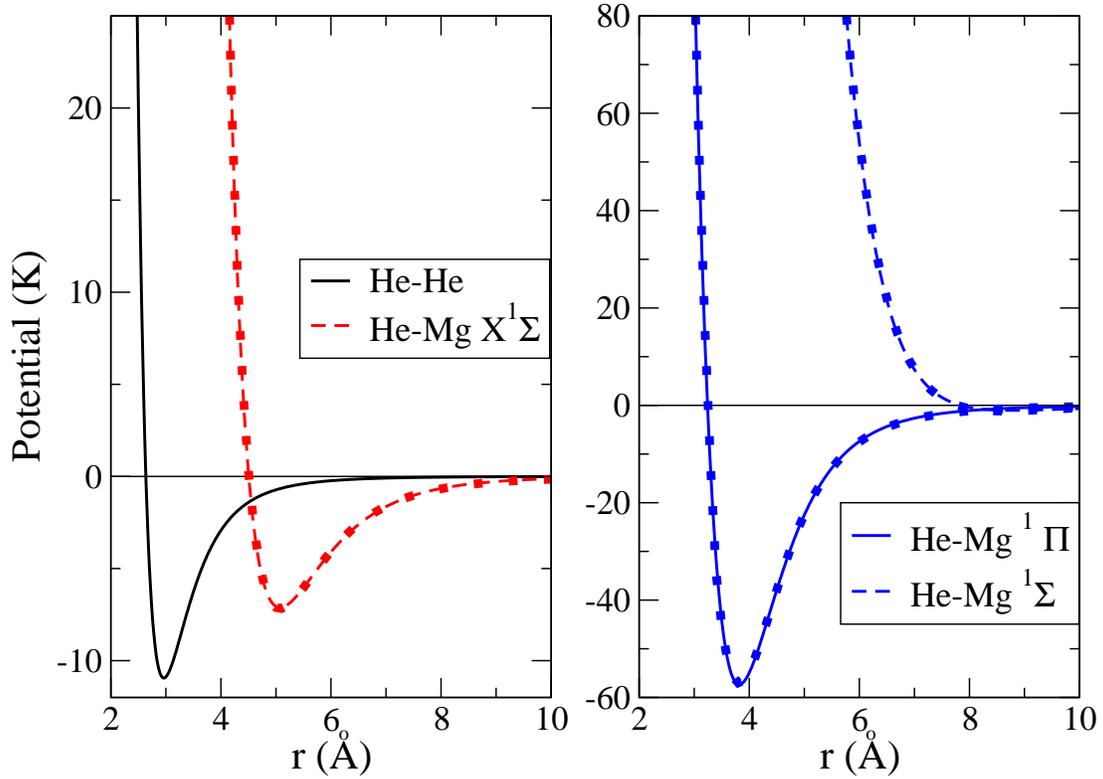}
\caption{(Color online) Pair potentials used in this work. The symbols correspond to
the results of Hinde \cite{Hin03} for the ground state (left panel) and Mella {\it et
al.}\cite{Mel05} for the excited states (right panel). The lines correspond to
the parameterizations given in Table \ref{params}. The Aziz He-He potential
\cite{Azi87} has also been plotted in the left panel (solid line).}
\label{pots}
\end{figure}

Quantum Monte Carlo methods aim at solving the Schr\"odinger equation of a
many-body 
system with the only knowledge of the inter-particle interaction.  
In our calculations we have employed the He-He Aziz potential,\cite{Azi87}
and the 
$X ^1\Sigma$ Mg-He interaction as obtained by Hinde.\cite{Hin03}
For the description of the absorption spectrum described in Sec \ref{spectra},
the 
$\Sigma$ and $\Pi$ Mg-He pair potentials of Ref. \onlinecite{Mel05} have been
used. 
To facilitate the use of these potentials, we have fitted them (by simulated
annealing) to an 
analytical expression of the kind 
\begin{equation}
\label{eq0}
V_{Mg-He}(r)= A e^{-\alpha r - \beta r^2} -
F(r)\times\left(\frac{C_6}{r^6}+\frac{C_{8}}{r^{8}} 
+\frac{C_{10}}{r^{10}}\right)
\end{equation}
$$
F(r) = \bigg\{
\begin{array}{lr}
 e^{-(1-D/r)^2} & r<D\\
 1 & r \geq D\\
\end{array}
$$
The parameters are given in Table \ref{params}. 
The resulting fits are shown in Fig. \ref{pots}, together with the 
calculated values (indicated with symbols in the figure). It can be seen that the
agreement is excellent. For the sake of  comparison, we have also
included the He-He interaction used in this work.  

Our DMC calculations are based on a variational or importance sampling wave
function. 
In that case, one does not solve the Schr\"odinger equation for the true wave
function 
$\Psi({\cal R},t)$, but for the auxiliary function $f({\cal R},t)= \Psi_T({\cal
R}) \Psi({\cal R},t)$. 
The trial wave function $\Psi_T({\cal R})$ guides the random walk and minimizes
the variance. 
We have used a rather simple form, containing the basic required properties, the
same as employed in the past to describe a Ca impurity.\cite{Gua09} It is a
generalization of the trial 
function adopted in previous studies on pristine mixed helium
clusters,\cite{Gua00a,Gua02}  
and is written as a product of seven terms
\begin{equation}
\Psi_T({\bf{\cal R}}) = \Psi_{44} \, \Psi_{33} \Psi_{34} \Psi_{4{\rm Mg}}
\Psi_{3{\rm Mg}}
D_{\uparrow} D_{\downarrow} \, ,
\label{function}
\end{equation}
where $\{ {\bf{\cal R}} \}$ represents the set of $3(N_3+N_4+1)$ coordinates of
the atoms forming the cluster,
$N_3 (N_4)$ being the number of $^3$He ($^4$He) atoms.
The first five terms are Jastrow factors $\Psi_{MN}$ for each pair $(M,N)$ of
different atoms, 
for which we have chosen the generic form
\begin{equation}
\Psi_{MN} = \prod_{i  \neq j} \exp\left( -\frac{1}{2}
\left[\frac{b_{MN}}{r_{ij}}\right]^{\nu_{MN}} -
\alpha_{MN} r_{ij} \right)  \; ,
\label{bosons}
\end{equation}
where indices $i,j$ run over the corresponding type of atom, and includes a
short-range 
repulsion term with parameters $b_{MN}$ and $\nu_{MN}$ associated with
it, and a long-range
 confining term with corresponding parameter $\alpha _{MN}$. 
The form of the short-range repulsion term was introduced long ago by 
McMillan~\cite{Mcm65}  to describe the homogeneous liquid $^4$He using
a 12-6 Lennard-Jones interaction. In that case, the values of the two parameters
$\nu$ 
and $b$ are fixed by the short-range behavior of a pair of atoms, and one gets
$\nu=5$, 
and $b=[16 \mu \epsilon/(25 \hbar^2)]^{1/10} \sigma^{6/5}$, where
$\mu$ is the reduced 
mass for each pair of atoms, and $\epsilon$ and $\sigma$ are the Lennard-Jones
energy and 
distance parameters, respectively. 

Notice that each $\Psi_{MN}$ function is explicitly symmetric under the exchange
of identical particles. 
The antisymmetry required for $^3$He fermions is incorporated in the Slater
determinants $D_{\uparrow}$ and $D_{\downarrow}$, related to the spin-up and -down fermions, respectively.
Two aspects should be considered in these determinants, namely the form and the filling of the single particle orbitals. 
As in previous works\cite{Gua00a,Gua02} the determinants have been built up with homogeneous monomials
of the fermion Cartesian coordinates as $x_i^{n_x} y_i^{n_y} z_i^{n_z}$, where the subindex $i$ refers to the particle, 
and the integer number  $n=n_x+ n_y+ n_z$ defines a ``shell". 
This kind of Slater determinant has
also been employed for other many-body systems, (see {\it e.g.} Ref.
\onlinecite{Lip03,Mah00}).
It turns out that they are of Vandermonde type and,
provided that the shells are filled in an increasing order of $n$, they
can be expressed in terms
of products of the relative coordinates, being thus translationally
invariant.\cite{Lip03,Mah00}
Besides, we have always assumed a filling scheme in which the total spin
is minimum, either 0 or 1/2, for $N_3$ even or odd, respectively.
For some pure $^3$He clusters, the shell filling has been given in
Ref. \onlinecite{Gua02}, indicating the total value of $S_z$. 

For the DMC algorithm we have used the  short-time Green function approximation
\cite{And80,Rey82} with an $O(\tau^3)$ form.\cite{Vrb86}
For those systems involving nodal surfaces, {\it i.e.} for fermions,
the fixed node approximation has been employed.\cite{And75,And76,Rey82}
The energy accuracy depends on the quality of the nodal surfaces of the
trial function, which arise from
the Slater determinants. Feynman-Cohen back-flow
correlations\cite{Fey56} have been incorporated into the scheme by
substituting 
\begin{equation}
{\bf r}_i \rightarrow {\tilde{\bf r}}_i =  {\bf r}_i + \sum_{i \ne j}
\eta(r_{ij})
({\bf r}_i - {\bf r}_j) 
\label{backflow}
\end{equation}
in the orbitals of the Slater determinants.\cite{Sch81} For the backflow
function $\eta(r)$ we choose the 
medium-range form used in Ref.~\onlinecite{Pan86}, namely
$\eta(r) = \lambda/r^3$, with the same value of $\lambda= 5$~\AA$^3$.
These type of correlations give rise to nodal surfaces that provide very
accurate results for different fermionic systems in Quantum Monte Carlo calculations.
\cite{Gua02,Cas00,Fou01,Bru03,Lop06,Par11,Sol06} 
Pure estimator results discussed later on are free of the trial wave function bias in
the density distributions. The accuracy of the wave function only affects the
rate of the convergence of the calculation. 

As in our previous works on pure and doped mixed helium
clusters,\cite{Gua09,Gua00a,Gua02} 
we have fixed the parameters $b_{MN}$ by taking $\epsilon$ as the minimum of the
interaction, and $\sigma$ as the distance at which the interaction is zero. 
Moreover, we have slightly 
modified the value of the exponent $\nu$ with respect to the McMillan value. 
All in all, for all cluster sizes we have used the 
following values:  $\nu_{MN}=5.2$, $b_{44}=2.95$~\AA,  $b_{33}=2.85$~\AA, 
$b_{34}= 2.90$~\AA, $b_{4{\rm Mg}}=5.831$~\AA, and $b_{3{\rm
Mg}}=5.687$~\AA. 
Thus, the trial or importance sampling wave function contains only five free
parameters, namely 
$\alpha_{44}$, $\alpha_{33}$, $\alpha_{34}$, $\alpha_{4{\rm Mg}}$, and
$\alpha_{3{\rm Mg}}$, 
which have been determined by minimizing the expectation value of the
Hamiltonian.
It turns out that for isotopically pure boson ($N_3=0$)
or fermion ($N_4=0$) clusters, these parameters are well
represented in terms of the number of atoms in the drop, namely
$\alpha_{44}= 0.117 - 0.029 N_4^{1/3}$, $\alpha_{4 \rm Mg}= 0.268 - 0.011 N_4^{1/3}$, 
$\alpha_{33}= 0.020 + 0.006 N_3^{1/3}$,
and $\alpha_{3 \rm Mg}= 0.191 + 0407 N_3^{1/3}-0.152 N_3^{2/3}$.

Our DMC calculation is based on an importance sampling function, and therefore the walkers 
are generated according to the auxiliary function 
$f({\cal R},t)= \Psi_T({\cal R}) \Psi({\cal R},t)$. 
The natural output corresponds thus to the so called mixed estimator, in which the
expectation value of a given operator is straightforwardly calculated with this 
probability distribution function.
If such an operator commutes with the Hamiltonian, the mixed estimator is equal to
the exact expectation value of the observable in the asymptotic limit and within the
fixed-node error. 
For other observables, as for example radial operators, the mixed estimator is
in general biased by the trial function used for importance sampling.
For the reasons mentioned in the Introduction, we have chosen to
obtain unbiased estimates of radial distances for some $^4$He droplets.
In view of the few unbiased available calculations,
ours may contribute to the general discussion of the subject.

Within Quantum Monte Carlo simulations, several schemes have been proposed in the 
literature to obtain 
unbiased---also called pure---estimators of operators that do not commute with the 
Hamiltonian.
In this work we compute pure expectation values by using a forward
walking method. The idea is to recover the exact value by including the factor 
$\Psi({\cal R},t)/\Psi_T({\cal R})$ in the expectation value.
This quotient can be obtained starting from the asymptotic offspring of the
walker.
\cite{Liu74} Therefore a weight proportional to the number of the future
descendant of the walker
\begin{equation}
W({\cal R})\propto n({\cal R}, t\rightarrow \infty)
\label{eq.descendant}
\end{equation}
needs to be included in the calculation.
We have employed the algorithm devised in Ref. \onlinecite{Bor95} to
calculate
this weight.
It makes use of an auxiliary variable associated with each walker that
evolves
with it, {\it i.e.} it is replicated as many times as the walker and
propagates the
local values. With the proper boundary conditions, the final average provides
the pure expectation value.
The basis of the method and details on the algorithm can be found
in Refs. \onlinecite{Bor95,sbc02}.
The pure estimation depends on the size of the block, $\Delta_L$, that
needs to be large enough to fulfill the forward walking condition Eq.
(\ref{eq.descendant}).
A study of the convergence as a function of $\Delta_L$ is
required to fix the block length from
which the pure estimator provides the same value within the statistical error
for a given radial operator.

\section{Results}
\label{results}

\subsection{Structure of Mg@$^4$He$_{N_4}$+$^3$He$_{N_3}$ clusters}

\begin{figure}[ht]
\begin{minipage}{.45\linewidth}
 \includegraphics[width=\linewidth,clip]{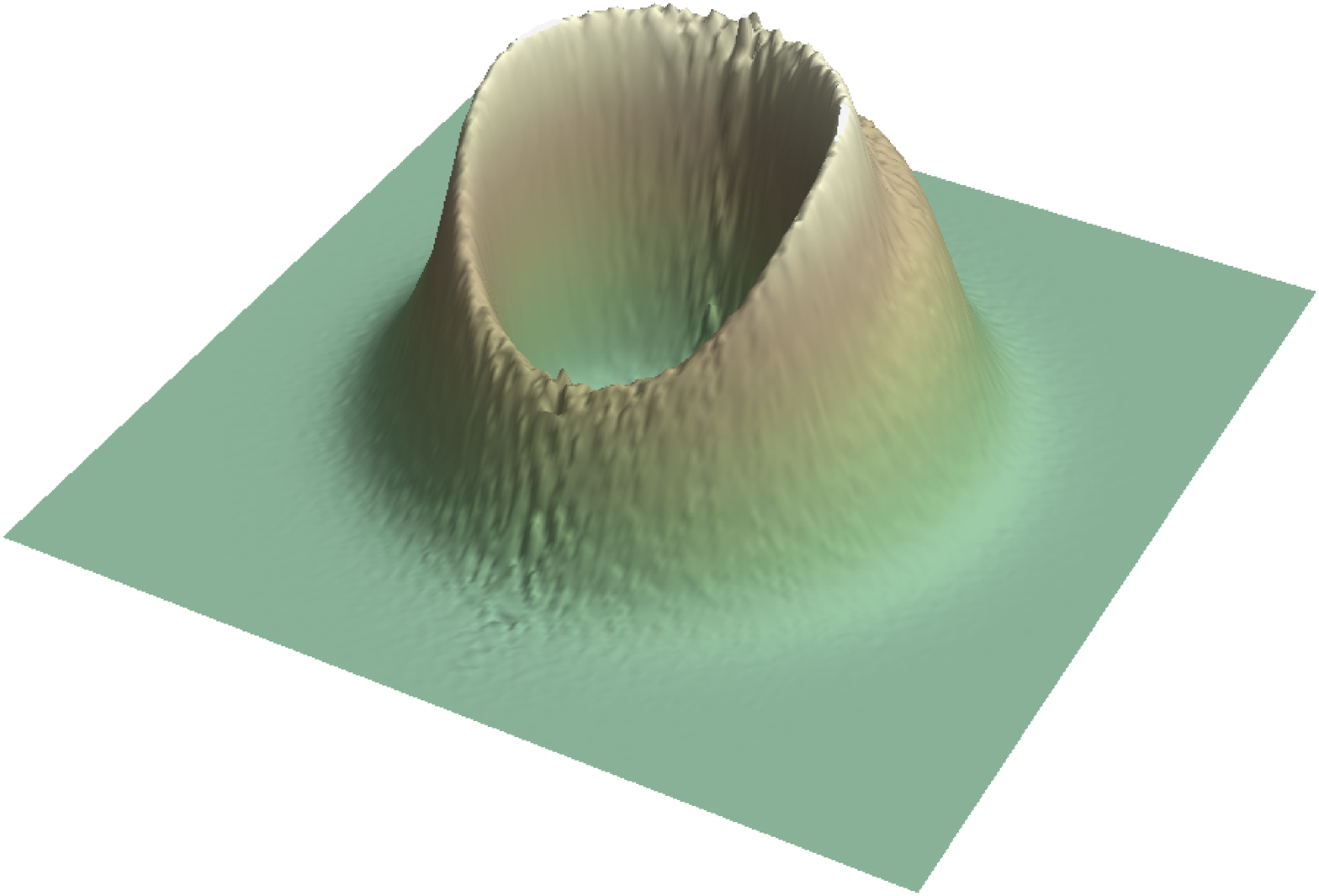}
 \includegraphics[width=\linewidth,clip]{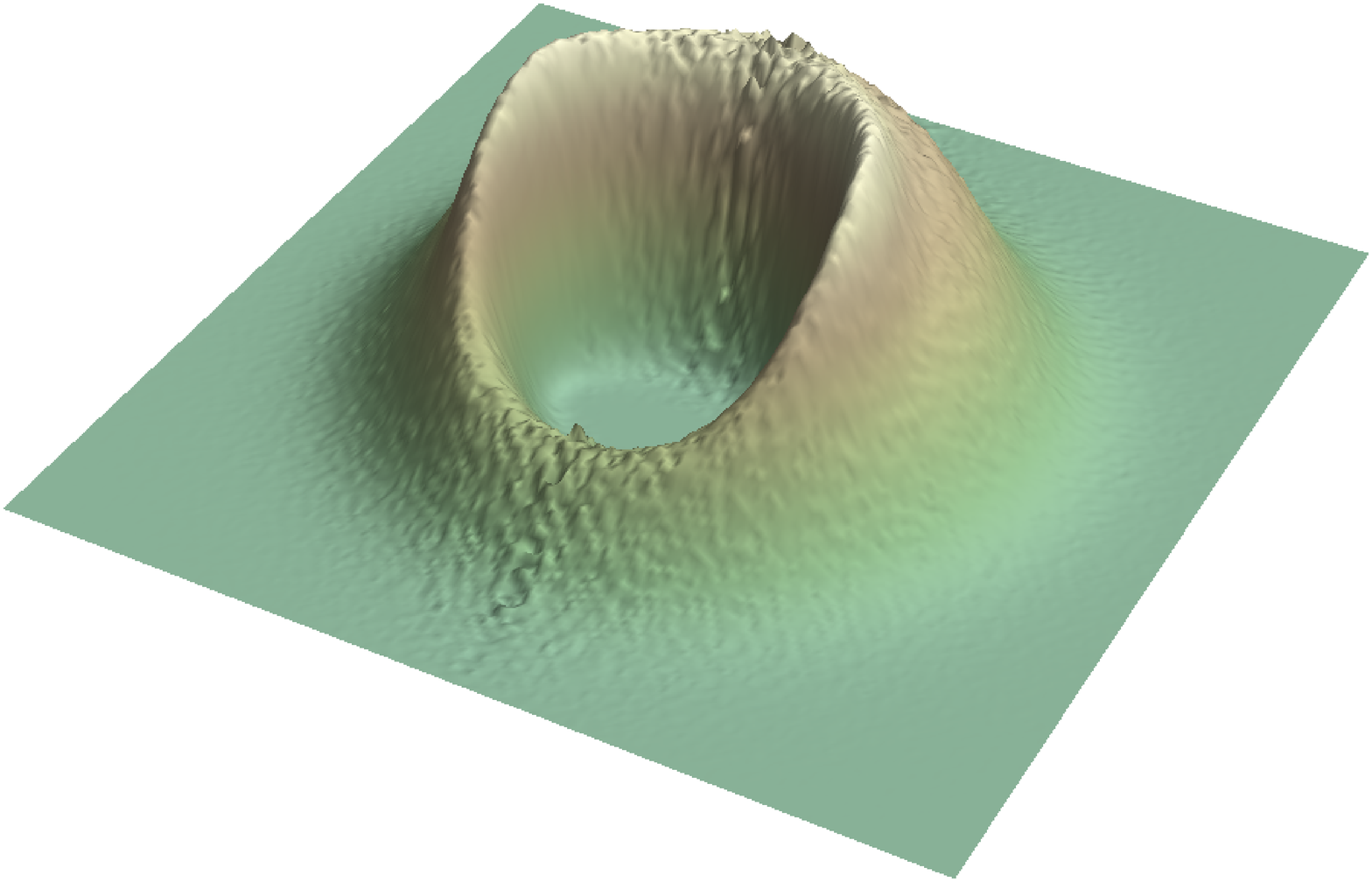}
 \includegraphics[width=\linewidth,clip]{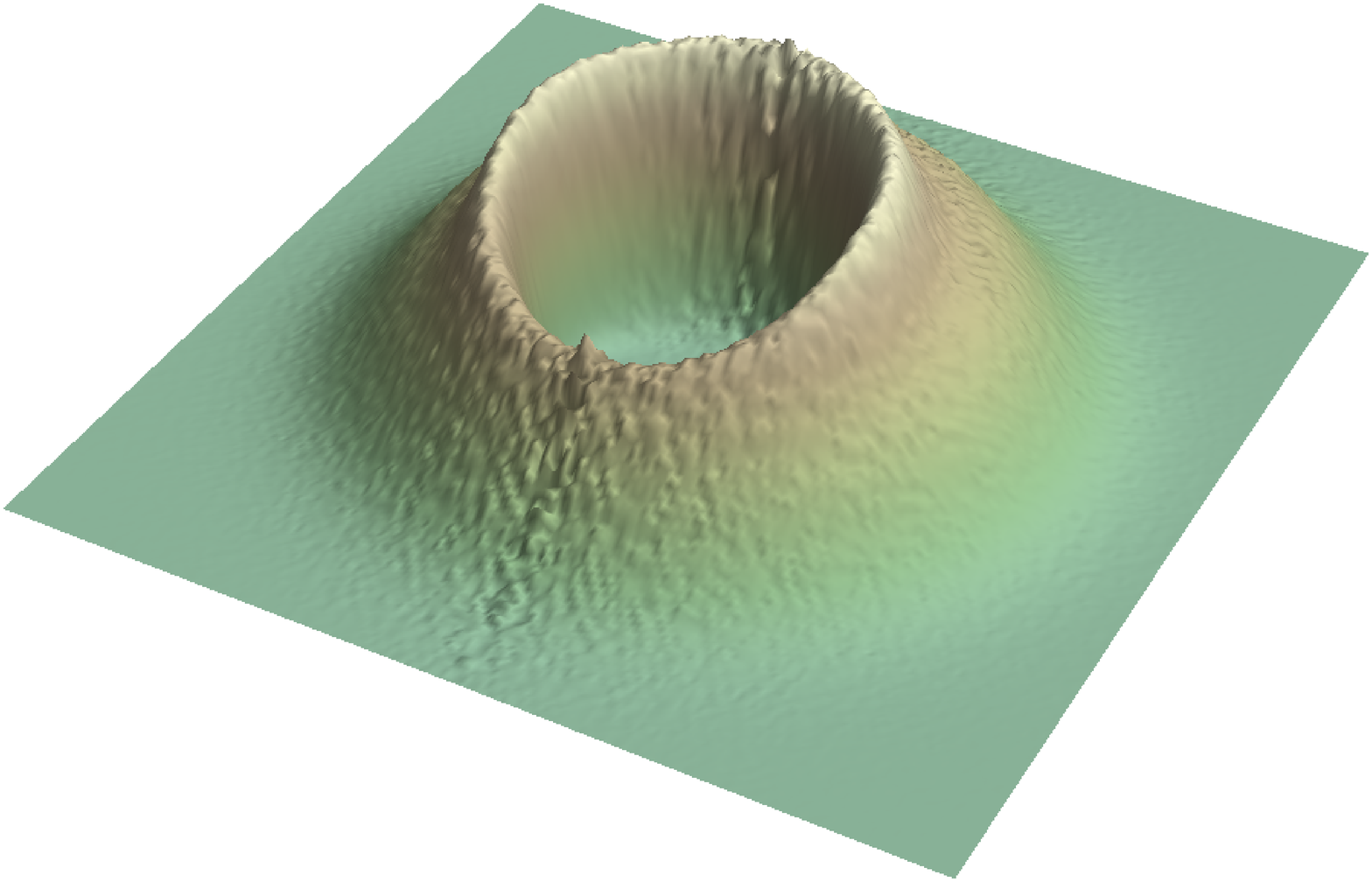}
\end{minipage}
\caption{(Color online)
Three-dimensional views of the total He atom density for
three droplets doped with Mg. From top to bottom:
$^4$He$_{40}$, $^3$He$_{20}$+$^4$He$_{20}$, and $^3$He$_{40}$.
}
\label{3Dplots}
\end{figure}

For illustrative purposes, the mixed estimator total helium
particle densities of the
$^4$He$_{40}$, $^3$He$_{20}$+$^4$He$_{20}$ and $^3$He$_{40}$ droplets 
doped with Mg are plotted in Fig. \ref{3Dplots}. These densities have been
obtained as follows.\cite{Gua09}
After a simulation running for a long thermalization time, we have
stored a large number of walkers (typically $10^6$) and for each of them
the origin is taken at the center of mass of the helium atoms. Next,
a rotation is carried out so that the Mg atom lies on the z-axis.
In this coordinate system, a projection onto the $y=0$ plane is performed in order to
compute the density.

It can be qualitatively
seen that, while for $^3$He$_{40}$ Mg is fully immersed, for
$^4$He$_{40}$ it is not. We will see in Sec \ref{spectra} how the
distinct helium environment around Mg is reflected in the absorption
spectrum for these three complexes.

\begin{figure}[ht]
\centering
\includegraphics[width=0.3\linewidth]{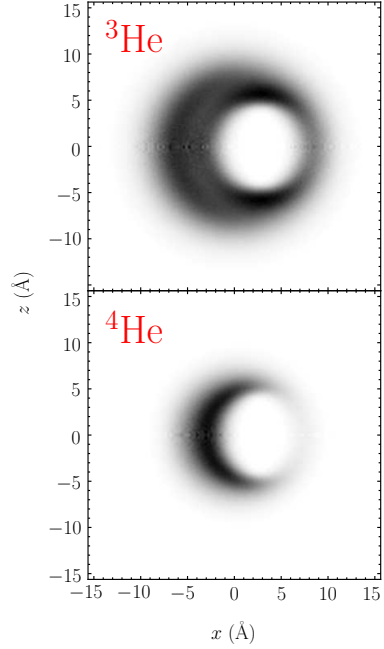}
\includegraphics[width=0.7\linewidth]{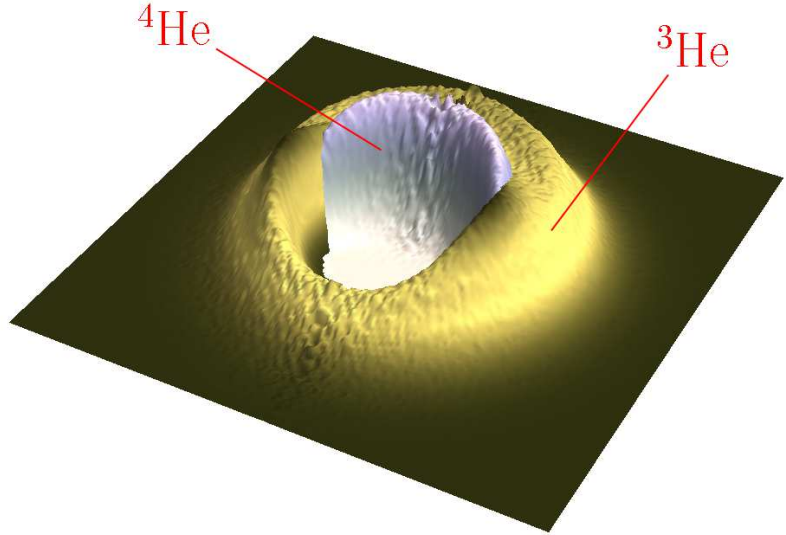}
\caption{(Color online) 
Atomic density of the mixed drop with $N_4=8$, $N_3=20$.
The contour plots of $^3$He and $^4$He densities are separately displayed
at the top of the figure. The darker the region, the higher the density.
The three-dimensional plot of the same densities is displayed at the bottom of 
the figure, with the same plot box as in the top contours.
}
\label{den0820}
\end{figure}

\begin{figure}[ht]
\centering
\includegraphics[width=0.3\linewidth]{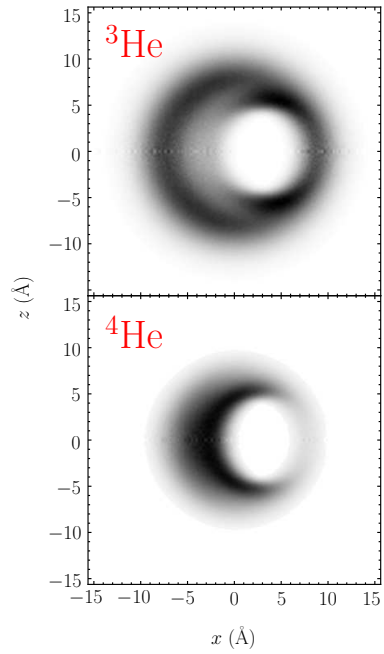}
\includegraphics[width=0.7\linewidth]{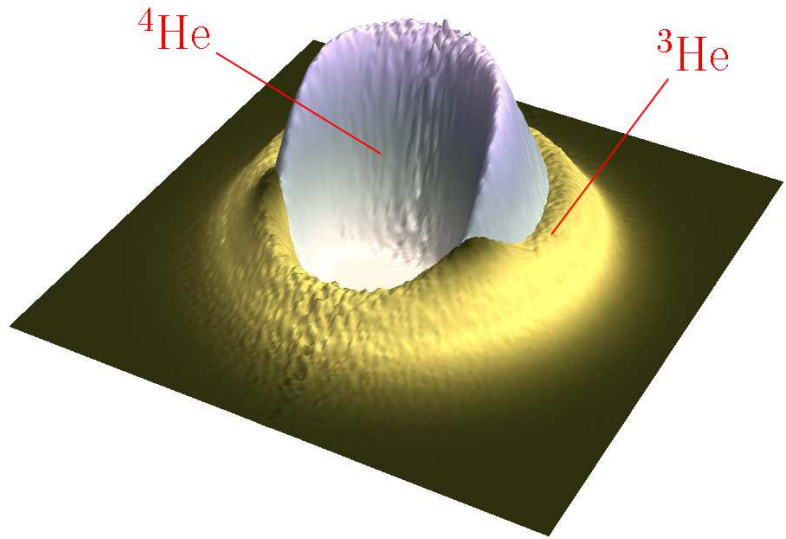}
\caption{(Color online) Same as Fig.~\ref{den0820} for $N_4=20,\, N_3=20$.}
\label{den2020}
\end{figure}

Figures \ref{den0820} and \ref{den2020} represent the mixed estimator
densities corresponding
to the doped droplets $^3$He$_8$+$^4$He$_{20}$ and
$^3$He$_{20}$+$^4$He$_{20}$, respectively. It is interesting to notice that
the $^3$He component
may get in touch with the impurity. This is even so for the $N_3= N_4 = 20$
system, the reason being that the balance between the weak He-He and Mg-He
interactions favors that the Mg atom is not fully
coated by $^4$He for such small droplets and the chosen composition.
The situation changes when $N_4$ increases.\cite{Gre98,Pi99}

\begin{figure}[ht]
\epsfig{file=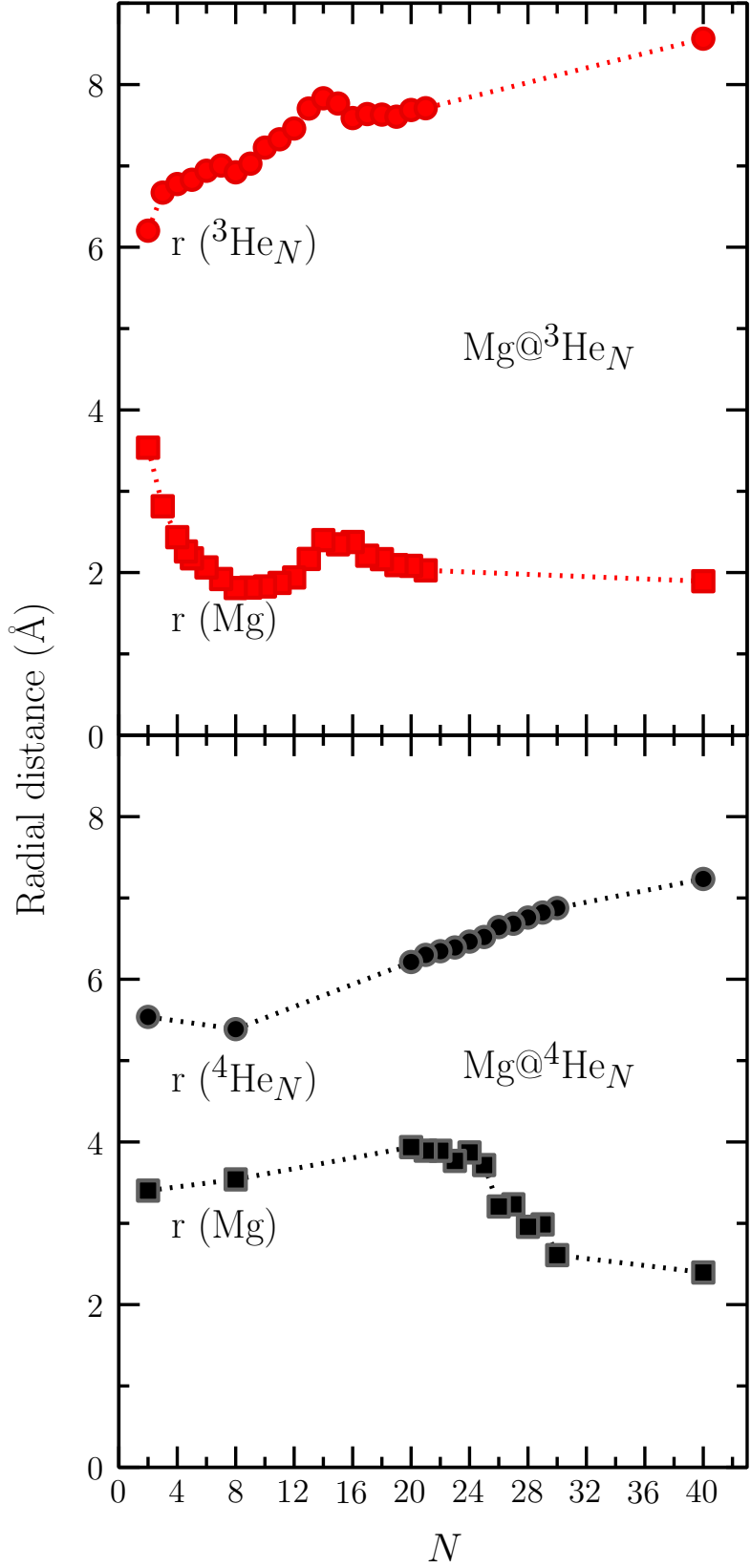,scale=1.6}
\caption{
Root mean square radii of the He clusters and the Mg impurity, measured with respect to 
the center-of-mass of the He cluster in both cases. 
The dotted lines are to guide the eye.}
\label{radiiCM}
\end{figure}

The mixed estimators
for the root mean square (rms) radius of the helium cluster and the Mg impurity are plotted in Fig.
\ref{radiiCM}.
We have chosen to refer them to the center-of-mass of the He droplet
instead of the center-of-mass of the He+Mg complex, so that
the data clearly show where the Mg sits with respect to the He moiety.

In $^4$He droplets the Mg rms radius increases with cluster size for small
($N_4 \lesssim 25$) droplets. One may infer from this that, for small
$N_4$ values, the impurity sits in the outer region of the droplet.
At $N_4 \simeq 25$ the
impurity begins to ``sink'' into the cluster as its rms radius starts to decrease
with increasing cluster size. 
In $^3$He droplets no such trend is seen for the Mg rms radius, which
contrarily has a tendency
to decrease as the cluster size increases, apart from the structure around $N_3=14$
for
whose origin, likely related to shell effects, we have been unable to find a
convincing explanation.
These results are consistent with the finding that Mg is in the bulk
of the droplet for any $N_3$ value,\cite{Her07} as Ancilotto's criterion
predicts.

\begin{figure}[ht]
\epsfig{file=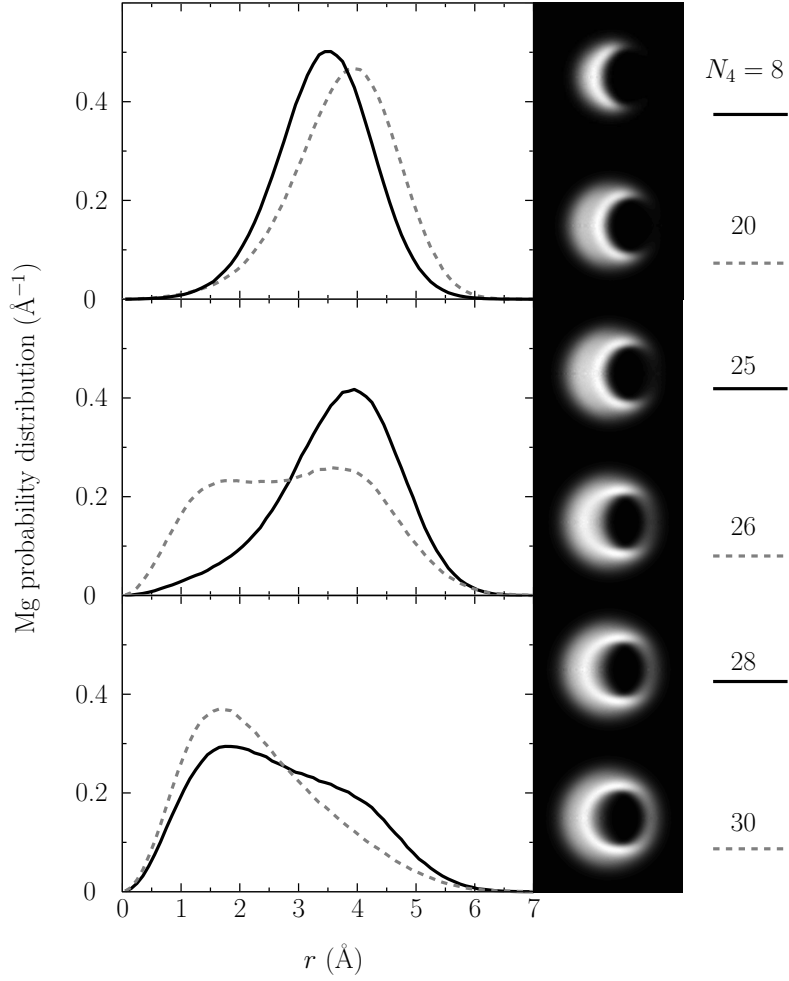,scale=0.6}
\caption{
Grey scale two-dimensional
plot of the particle density (right panel)
for different Mg@$^4$He$_{N_4}$ clusters; the lighter the region, the higher the
helium density. 
The He densities are displayed in boxes of 30 \AA \, $\times$ 30 \AA.
The left panels represent
the probability distribution $P(r)$ 
of finding the Mg atom at a distance $r$  from the He center-of-mass
\big[$\int_0^\infty dr P(r) = 1$\big].
}
\label{densites}
\end{figure}

The transition in the Mg location Fig.~\ref{radiiCM} hints at can be clearly 
seen in Fig.~\ref{densites}, where the mixed estimator radial probability
distribution of Mg in $^4$He clusters is shown for several sizes from
$N_4=8$ to 30.
For small cluster sizes ($N_4 < 25$), the Mg atom is found at the surface of the
droplet. The typical distance between the impurity and the center-of-mass of the
droplet is about 3-4 \AA. At $N_4 = 25$ Mg experiences a transition from a surface
to a bulk state in which the probability of finding the impurity inside
the droplet becomes significant. For cluster sizes
between $N_4=25$ and 29, the impurity is highly delocalized, and for
$N_4 \ge 30$
the Mg atom resides inside the droplet. The typical distance between the
impurity and
the center-of-mass of the droplet decreases as the impurity becomes fully solvated.
Our results thus confirm those of Mella {\it et al.}\cite{Mel05}

\begin{figure}[ht]
\epsfig{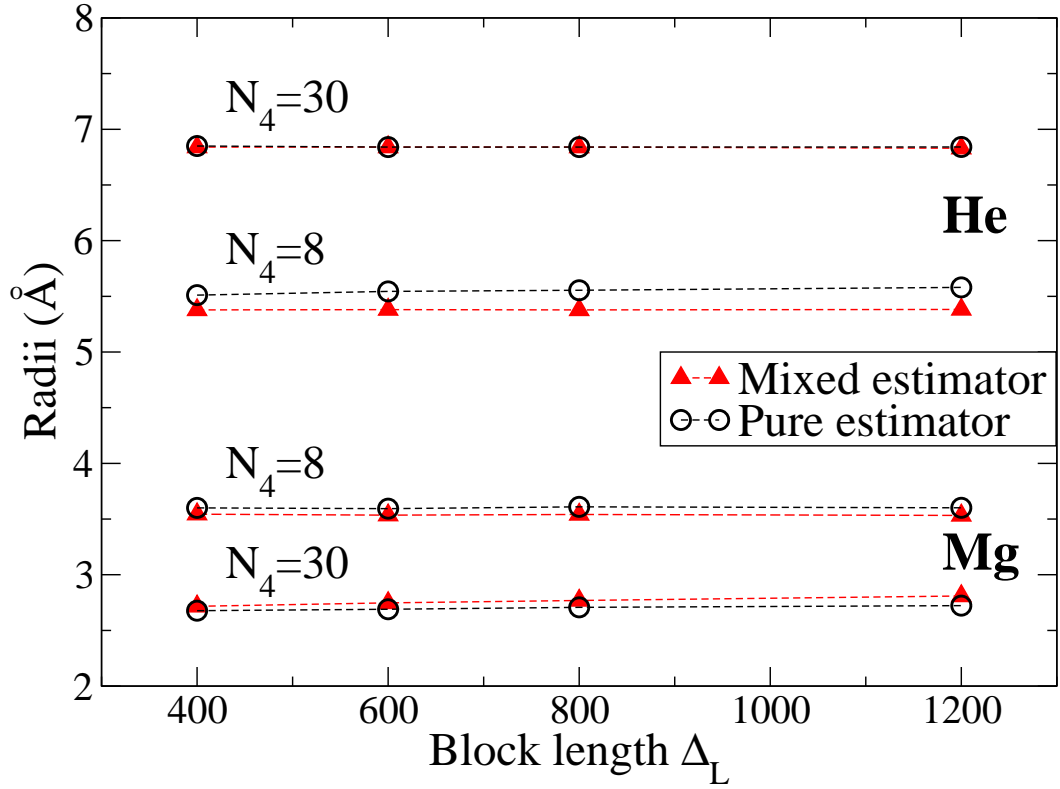}
\caption{(Color online) 
Evolution of the rms radii with the length of blocks of the pure estimator.
The dashed lines are to guide the eye.}
\label{radiiMP}
\end{figure}

Finally, we have employed the {\it pure estimator} to determine the
radial structure for the isotopically pure $N_4=8$ and 30 droplets.
We have found that the differences between pure and mixed estimators of the radial densities 
are very small and not worth to be plotted. 
As an illustration, we have shown in Fig. \ref{radiiMP} the mixed and pure estimations for the 
rms radius as a function of the block size $\Delta_L$.
The stability of the results is apparent, as well as the quality of our mixed
estimator for radial distances. Thus, the conclusions drawn
using the mixed estimator are robust and remain unchanged and
accordingly, 
apart from Fig. \ref{radiiMP}, the results discussed in
this work have all been obtained using the mixed estimator.

\subsection{Energetics of Mg@$^4$He$_{N_4}$+$^3$He$_{N_3}$ clusters}

The calculated DMC ground state energies $E({\rm Mg}@^4{\rm He}_{N_4}+
^3{\rm He}_{N_3})$ are given in Tables \ref{ener4}, \ref{ener34},
and \ref{ener3}  for several $(N_3, N_4)$ combinations. 
In these Tables we have shown the statistical error in the last figure obtained in the
usual way, as the standard deviation of the mean value calculated by using the
blocking method.
The solvation energy of the dopant in the droplet, defined as
\begin{equation}
\mu_{\rm Mg} = E({\rm Mg}@^4{\rm He}_{N_4}+^3{\rm He}_{N_3})
- E(^4{\rm He}_{N_4}+^3{\rm He}_{N_3})\, ,
\end{equation} 
is also given in the Tables.
No solvation energy is given for Mg$@^3{\rm He}_{N_3}$ droplets
because pure $^3$He droplets are unbound for such
small sizes\cite{Gua00a,Sol06,Bar97b} and hence the solvation
energy
is just the ground state energy of the Mg$@^3{\rm He}_{N_3}$ complex.
At first glance, it is surprising that one single Mg atom is able
to bound any number of $^3$He atoms, since the Mg-He interaction is
weaker than the He-He one, see Fig.~\ref{pots}. 
The smaller zero-point energy of Mg together with the fact that the
equilibrium distance is larger for Mg-He than for He-He causes
the extra binding.
Thus, the complexes Mg@$^4$He$_{N_4}$, Mg@$^3$He$_{N_3}$ and 
Mg@$^4$He$_{N_4}$+$^3$He$_{N_3}$ are bound for any $N_3$ and $N_4$ values.

A comparison between our DMC results for Mg@$^4$He$_{N_4}$ clusters and
other DMC results from Refs. \onlinecite{Mel05} and \onlinecite{Elh09}
is presented in Table \ref{compa4}. Mella {\it et al.}
\cite{Mel05} have used a Mg-He interaction determined at the CCSDT
level, and the He-He interaction of Ref. \onlinecite{Tan95}. 
The differences beyond statistical errors found between our results and
theirs could be mostly attributed to minor differences in the
interaction potentials. The agreement with
Elhyiani\cite{Elh09} -who has used the same
interactions as in this work- is very satisfactory.

\begin{figure}[ht]
\epsfig{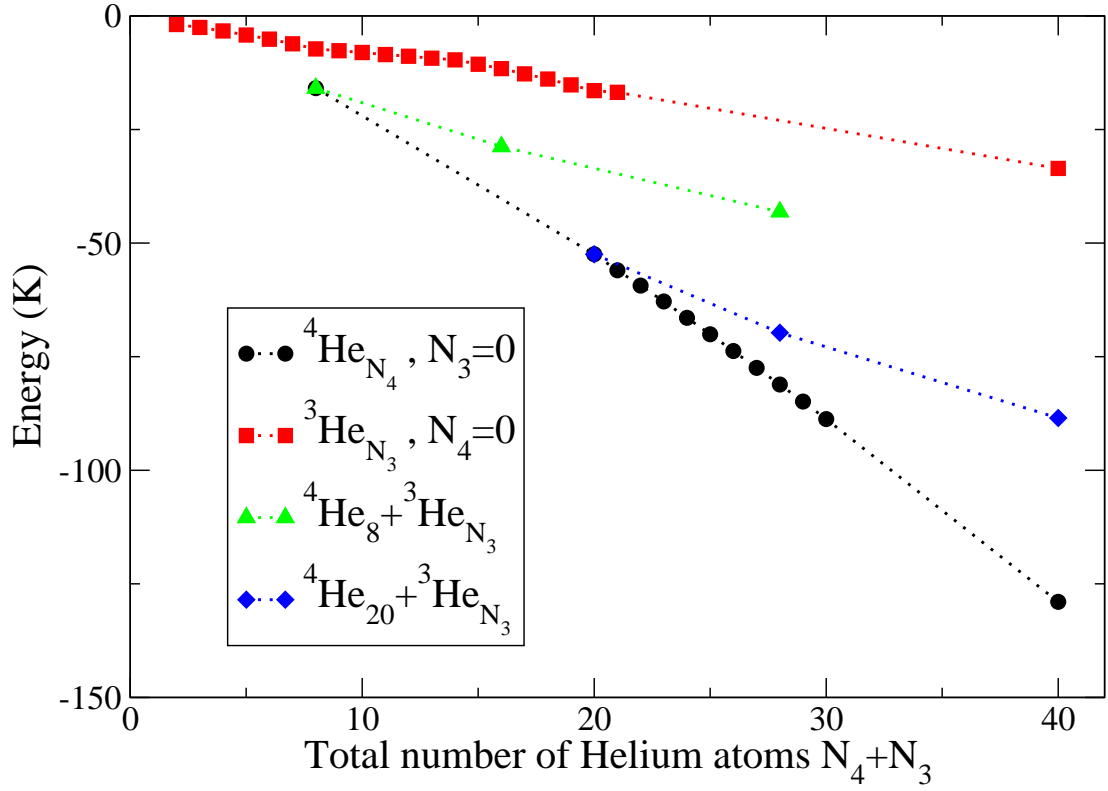}
\caption{(Color online) 
Ground state total energies (in K) of Mg@$^4$He$_{N_4}$+$^3$He$_{N_3}$ droplets. 
The dotted lines are to guide the eye.
}
\label{ener}
\end{figure}

\begin{figure}[ht]
\epsfig{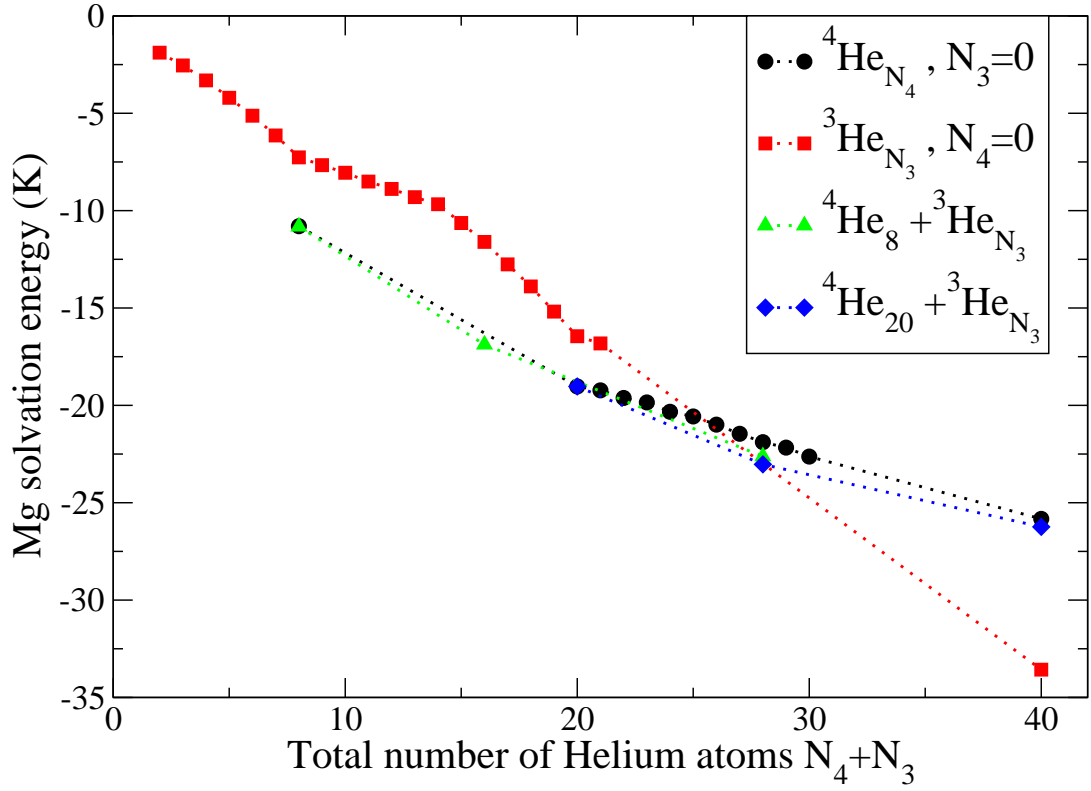}
\caption{(Color online)
Solvation energy of Mg (in K) in Mg@$^4$He$_{N_4}$+$^3$He$_{N_3}$ droplets. 
The dotted lines are to guide the eye.
}
\label{mgbind}
\end{figure}

The ground state and Mg solvation energies 
are presented in Figs. \ref{ener} and \ref{mgbind} respectively,
as a function of the total number of He atoms.
For the sizes considered here, it can be seen that the Mg solvation 
energy does not much change when $^3$He atoms are added to the
Mg@$^4$He$_{N_4}$ droplet.

\begin{figure}[ht]
\epsfig{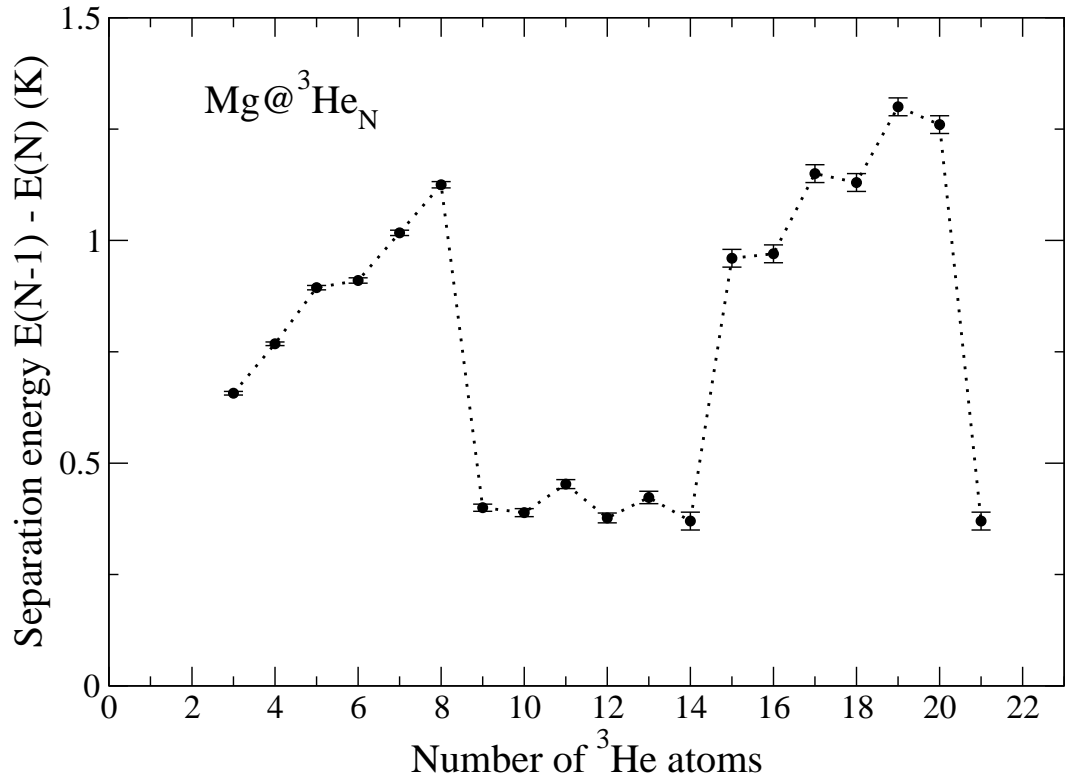}
\caption{
Separation energy of a $^3$He atom (in K) in Mg@$^3$He$_{N_3}$ droplets. 
The dotted line is to guide the eye.
}
\label{sepaHe3}
\end{figure}

Both ground state and solvation energies are smooth functions of the
total number of helium atoms. 
It is worthwhile noting that for doped $^3$He droplets, these energies
display conspicuous oscillations. 
We have defined the $^3$He separation energy as
\begin{equation}
S_{\rm ^3He} =
E({\rm Mg}@^3{\rm He}_{N_3-1}) -E({\rm Mg}@^3{\rm He}_{N_3}) \, ,
\end{equation} 
and have plotted this quantity as a function of $N_3$ in Fig. \ref{sepaHe3}.
$S_{\rm ^3He}$ has a sawtooth structure similar to that of the atomic
ionization energy {\it vs}. atomic number.\cite{Eis85} The more tightly bound
$^3$He atoms at
$N_3$=8 and 20 allows one to identify shell closures, and correlate
well with the  local minima in Figs. \ref{ener} and
\ref{mgbind}.
A similar result was obtained for
Ca.\cite{Gua09}
This is somewhat an unexpected result, as
$^3$He droplets doped with Mg atoms look closer to axially than
to spherically symmetric systems. 
We recall that for spin saturated systems,
the first shell closures of the
three-dimensional spherical harmonic oscillator appear for 
2, 8, and 20 spin 1/2 fermions.

\subsection{Absorption spectrum of Mg in$^4$He$_{N_4}$+$^3$He$_{N_3}$ clusters}
\label{spectra}
It is well-known that the shift and width of the electronic
transitions of impurities in helium droplets
are very sensitive to their environment and for this reason this technique
is often employed to study their structure.\cite{Sti06,Tig07}
We have calculated the dipole absorption spectrum of Mg
as described in Ref. \onlinecite{Gua09}.
The line shape of the electronic transition is determined as
\begin{equation}
I(\omega)  \propto \int {\rm d}{\bf {\cal R}} |\Psi_{gs}({\bf {\cal R}})|^2 
\delta( \omega+V_{gs}({\bf {\cal R}})-V_{ex}({\bf {\cal R}}) ) \;\; ,
\label{absorption}
\end{equation}
where $\{\bf {\cal R} \}$ refers to the positions of the atoms, and $V_{gs}$ and
$V_{ex}$ 
are, respectively, the ground and excited states potential energy surfaces.
We recall that we have used the Mg-He $X^1\Sigma$ interaction of
Ref.~\onlinecite{Hin03} for
the ground state and the $^1\Pi$ and $^1\Sigma$ potentials of
Ref.~\onlinecite{Mel05} for the excited states. To compute $I(\omega)$
for a given value of
$\omega$, one has to diagonalize a $3\times 3$ matrix to determine the three
components of the absorption line, each one arising from a different potential
energy surface, {\it i.e.}, 
eigenvalue of the excited energy matrix.
We refer the reader to Refs. \onlinecite{Che96,Nak01,Mel02,Her08b,Her09b}
and references therein for the details.

The DMC calculation provides us with a set of walkers $\{ {\cal R}_j \}$ 
representing the instantaneous position of each atom in the cluster. These walkers
have been used for determining the one-body  densities presented before
and can be also employed to obtain the absorption spectrum
replacing 
$\int {\rm d}{\bf {\cal R}} |\Psi_{gs}({\bf {\cal R}})|^2$ by a sum over
$\{ {\cal R}_j \}$ in the above equation, so that 
a mixed estimator of $I(\omega)$ has been obtained. 
This is the same approximation employed for the calculation of the radial densities
here studied finding no significant differences with their pure estimator values.

\begin{figure}[ht]
\epsfig{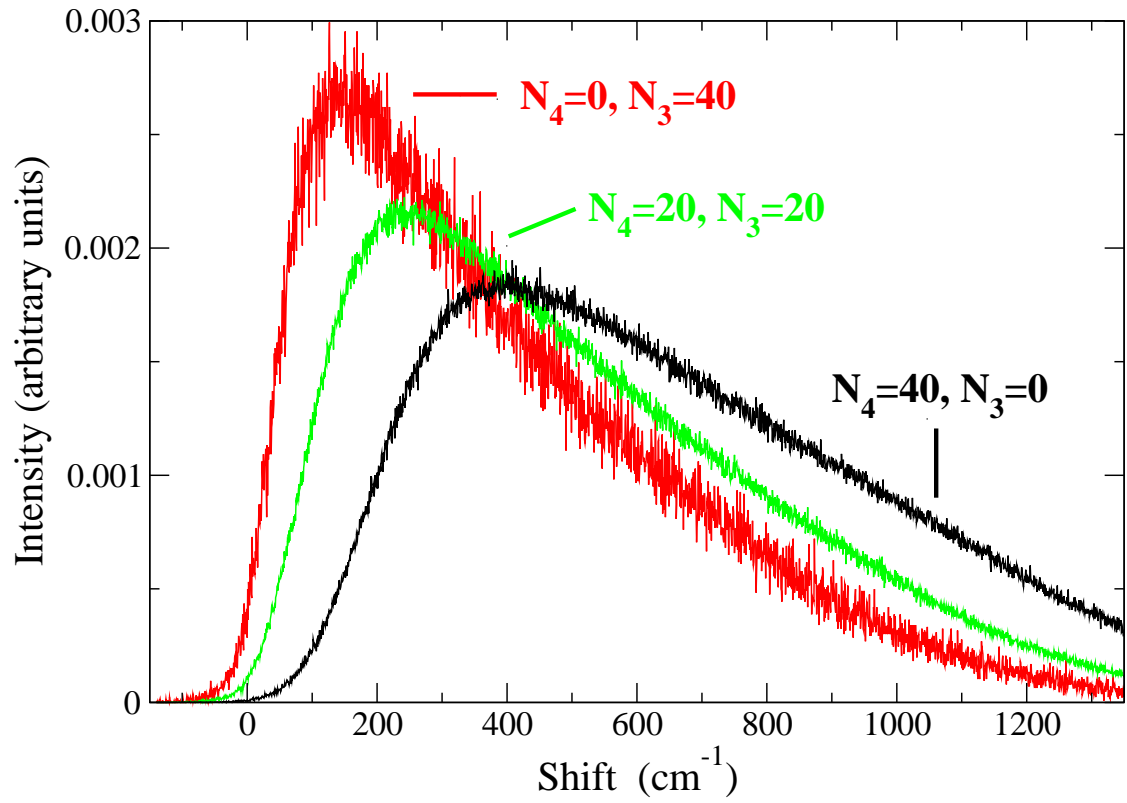}
\caption{(Color online) Dipole absorption lines of Mg in three selected $N_3+N_4=40$
droplets.} 
\label{spectra40gros}
\end{figure}

As an illustrative example, the absorption spectrum of Mg is plotted 
in Fig. \ref{spectra40gros} for three
selected $(N_3, N_4)$ combinations with $N_3+N_4=40$.
The spectra are referred to that of Mg in the gas phase.
As for other impurities, the long tail at high frequencies
arises from the very repulsive contribution of the $^1\Sigma$ pair potential. 
The largest atomic shift  corresponds to the $^4$He droplet and the smallest one 
to the $^3$He droplet, the one corresponding to the mixed droplet lies in between.

The relative value of the shifts
is easy to understand from the appearance of the helium densities
shown in Fig. \ref{3Dplots}.
Roughly speaking, the larger the density around the impurity, the larger the shift.
For $N_3+N_4=40$, Mg is fairly coated by helium and  the value of the
saturation density plays a significant role in the actual value of the shift.
Since the saturation density is larger for
$^4$He than for $^3$He this explains the relative position of the three absorption lines.

\section{Summary}
\label{conclusion}

Using a DMC approach, we have found that a Mg atom in a
$^4$He droplet experiences a transition from a surface to a bulk
location for $N_4 \ge 26$. 
For larger $^4$He droplets, the impurity resides in the bulk of the droplet. 
This conclusion has been drawn not only using the mixed estimator
inherent to the importance sampling approach, but also using a pure estimator
approach free from this bias. This finding agrees with the result obtained
by Mella and coworkers,\cite{Mel05} with which only minor quantitative
differences are found.

Due to the light mass of this alkaline earth atom and its weak 
interaction with helium, Mg impurities are found to
be very delocalized inside droplets containing several thousand $^4$He
atoms,\cite{Her08}
as those of interest in recent experiments.\cite{Prz08,Ren07}
In the case of $^3$He, we have found that Mg is always in the
bulk of the $^3$He droplet, as determined by previous DFT
calculations.\cite{Her07} To the best of our knowledge, the present
calculations are the only DMC ones available for this system and for
isotopically mixed helium droplets doped with Mg as well.

For mixed droplets, we have found the well-known
scenario\cite{Gre98,Pi99} that $^3$He and $^4$He atoms are
distributed around the impurity into a onion-like shell structure,
with $^4$He atoms coating the Mg impurity. This happens once the number of
$^4$He atoms is large enough to fully cover the dopant. For small $N_4$ values,
it may appear that some $^3$He is in contact with the Mg atom, which is
again a consequence of the surface location of Mg in small $^4$He clusters.
We want to point out, however, that distinguishing surface from volume regions
in such small systems is largely arbitrary.

The DMC walkers have been employed to calculate the dipole absorption line
taking as a case of study three selected clusters with $N_3+N_4=40$. We have 
found that the atomic shift is different enough for the studied configurations.
This result only pertains to small mixed droplets. Indeed,
when the number of $^4$He atoms in the droplet is large enough, the shift
will be insensitive to the actual composition of the droplet, as the $^3$He
component will be distributed into a shell distant from the impurity, whose
absorption line will quickly tend to that of Mg in liquid 
$^4$He.\cite{Mor99,Mor06}

\acknowledgments

This work has been supported by Grants FIS2007-60133, 
FIS2008-00421/FIS and FIS2009-07390
from DGI, Spain (FEDER), 2009SGR1289 from Generalitat de
Catalunya, and by the Junta de Andaluc\'{\i}a.
DM has been supported by the ME (Spain) FPU program, Grant
No. AP2008-04343.

\newpage

\begin{table}[htb]
\caption{Parameters of the fit to the calculated Mg-He pair potentials. The
values are such that Eq.~(\ref{eq0}) yields the pair potential in K when the distance $r$ is expressed in \AA{}.}
\begin{tabular}{ccccccc} 
\hline
        && $X ^1\Sigma$ (Ref. \onlinecite{Hin03}) && $^1\Pi$ (Ref.
\onlinecite{Mel05}) && $^1\Sigma$  (Ref. \onlinecite{Mel05})   \\
\hline 
$A$   && $1.16902 \times 10^{7}$  && $1.1308 \times 10^{5}$   && $8.6195 \times
10^{3}$ \\
$\alpha$ && $3.0188$   && $1.6555$       && $0.00291$ \\
$\beta$  && $-$              && $0.17578$     && $0.14011$  \\
$D$     && $10.002$        && $9.3539$       && $ 21.075$  \\
$C_6$  && $1.9454 \times 10^{4}$ && $2.0241 \times 10^{5}$ &&$-$                
    \\
$C_8$   && $1.04717 \times 10^{7}$  && $-$    && $-$      \\
$C_{10}$   && $-$  && $3.5046 \times 10^{8}$ && $2.3161 \times 10^{10}$ \\
\hline  
\end{tabular}
\label{params}
\end{table}

\newpage 

\begin{table}[ht]
\caption{Ground state energies of Mg@$^4$He$_{N_4}$ and Mg solvation energy
(in K). }
\begin{center}
\begin{tabular}{ccc}
\hline 
$N_4$ & Ground state energy & $\mu_{\rm Mg}$   \\
\hline
 8   &  -15.91   (1)   &  -10.80 (2)           \\
20  &  -52.46   (2)   & -18.52 (7)           \\
21  &  -55.99   (2)   &  -19.23 (5)             \\
22  &  -59.35   (2)   &  -19.62 (6)      \\
23  &  -62.84   (2)   &  -19.85 (7)       \\
24  &  -66.44   (2)   &  -20.33 (9)        \\
25  &  -70.07   (2)   &  -20.57 (8)        \\
26  &  -73.75   (2)   &  -20.99 (9)         \\
27  &  -77.46   (2)   &  -21.46 (9)         \\
28  &  -81.13   (2)   &  -21.89 (9)        \\
29  &  -84.87   (2)   &  -22.17 (7)         \\
30  &  -88.74   (2)   &  -22.63 (10)         \\
40  & -128.96  (3)   &  -25.83 (12)          \\
\hline 
\end{tabular}
\end{center}
\label{ener4}
\end{table}

\newpage

\begin{table}[ht]
\caption{Ground state energies of Mg@$^4$He$_{N_4}$+$^3$He$_{N_3}$ and
Mg solvation energy (in K).}
\begin{center}
\begin{tabular}{cccccc}
\hline 
$N_4$ & $N_3$ && Ground state energy && $\mu_{\rm Mg}$   \\
\hline
 2  & 2    &&   -5.312  (6) &&   -5.201 (6) \\ 
 8  & 8    &&  -28.77  (2)  &&   -16.89 (2)\\ 
20 & 8    &&  -69.73  (4)  &&   -23.03 (4) \\ 
 8 & 20   &&  -43.13 (2)   &&   -22.62 (3) \\ 
  20 &20 &&  -88.51 (5)   &&   -26.25 (6) \\ 
  \hline 
  \end{tabular}
\end{center}
\label{ener34}
\end{table}

\newpage 

\begin{table}[ht]
\caption{Ground state energies of Mg@$^3$He$_{N_3}$ (in K).}
\begin{center}
\begin{tabular}{cccc}
\hline 
$N_3$ & Ground state energy  & $N_3$ & Ground state energy  \\
\hline
 2   &  -1.887  (2) & 13   & -9.31   (1) \\
 3   &  -2.544  (3)  &  14   & -9.67   (1) \\ 
 4   &  -3.312  (3)  & 15  & -10.64   (1) \\ 
 5   &  -4.206   (4) & 16  & -11.61 (1)            \\
 6   &  -5.124  (4)  & 17  & -12.76  (1) \\ 
 7   &  -6.141  (5)  & 18  & -13.89   (1) \\ 
 8   &  -7.266   (5) &  19  & -15.19   (1) \\        
 9   & -7.666   (6)  & 20   &   -16.45   (1)         \\
10  &  -8.055  (7)  & 21    &  -16.820 (1)          \\
11  & -8.508   (7)  & 40   &   -33.57  (3)          \\
12   & -8.885   (8) &   &  \\ 
\hline 
\end{tabular}
\end{center}
\label{ener3}
\end{table}

\newpage

\begin{table}[ht]
\caption{Comparison of several DMC results for Mg@$^4$He$_{N_4}$ clusters.}
\begin{center}
\begin{tabular}{c|ccc|ccc}
\hline 
 & \multicolumn{3}{@{}|c|@{}}{Cluster energy (K)}& 
\multicolumn{3}{@{}|c@{}}{Mg solvation energy (K)} \\ 
\hline
$N_4$ & This work & Ref. \onlinecite{Mel05} & Ref. \onlinecite{Elh09} 
& This work & Ref. \onlinecite{Mel05} & Ref. \onlinecite{Elh09} \\
\hline
8   & -15.91 (1) &                  & -15.91 (1) & -10.80 (2) &                &
-10.76 (1) \\
20 & -52.46 (2) & -51.64 (1)  &                 & -18.52 (7) & -18.49 (1) &       
         \\
25 & -70.07 (2) & -69.04 (1)  & -70.15 (6) & -20.57 (8) & -25.03 (4) & -20.93 (9)
\\
30 & -88.74 (2) & -87.41 (1) &                   & -22.63 (10) & -26.89 (4) & \\
40 & -128.96 (3) & -126.51 (3) &               & -25.83 (12) & -25.39 (4) & \\
  \hline 
  \end{tabular}
\end{center}
\label{compa4}
\end{table}

\newpage

\newpage

\newpage

\newpage 

\newpage

\newpage

\newpage

\newpage

\newpage

\newpage

\newpage

\end{document}